\documentclass[a4paper,12pt]{article}
\usepackage{graphicx}
\newcommand{\DtoKpipi}{{\rm D^+\to K^-\pi^+\pi^+}}

\begin{document}

\markboth{Elena Bruna}{Open charm reconstruction in ALICE: $\DtoKpipi$}

\title{Open charm reconstruction in ALICE: $\DtoKpipi$
}

\author{Elena Bruna \\
for the ALICE Collaboration\\
{\small University of Torino} \\
{\small Via P. Giuria 1 10125, Torino, Italy}\\
{\small elena.bruna@to.infn.it}
}
\maketitle
\begin{abstract}
Open charm mesons produced in high energy A-A interactions are expected to be powerful probes to investigate the medium produced in the collision. 
In this context it is important to measure the production of as many charmed hadrons as possible, such as D$^0$, D$^+$, D$^+_s$ and $\Lambda_c$, because the measurement of their relative yield can provide information on the hadronization mechanism and is necessary to reduce the systematic error on the absolute cross section.
The ALICE experiment at the LHC is designed to perform such measurements 
at midrapidity down to $p_T$ below 1 GeV/c, mainly by means of the silicon 
vertex and tracker detector, the time projection chamber and the time 
of flight detector.
One of the main channels for the detection of charm production in ALICE is the exclusive reconstruction of the D$^+$ meson through
its three charged body decay $K^-\pi^+\pi^+$ in Pb-Pb ($\sqrt s=5.5$ TeV) and
pp ($\sqrt s=14$ TeV) collisions.
The selection strategies for this analysis and the results of a
feasibility study on Monte Carlo events will be presented together with the 
perspectives for the study of D$^+$ quenching and azimuthal anisotropy measurements.
\end{abstract}

\section{Heavy quarks in nucleus-nucleus collisions at high energy}\label{sec:HQprod}
Heavy-quark production takes place in partonic scatterings during the early stages of the nucleus-nucleus collision. The time scale for a $c\bar c$ pair production is \mbox{$\sim\hbar / (2m_Qc^2)\simeq (0.2$}GeV fm c$^{-1})/(2.4$GeV$) \simeq 0.1$ fm/c, which is much smaller than the expected lifetime of the Quark Gluon Plasma \mbox{$\sim10$} fm/c. Thus, heavy quarks are expected to provide information about the hottest initial phase.
The measurement of D mesons can be used to extract the charm production cross section. The measurement of charm cross section in both pp and AA collisions is useful to evaluate the scaling mechanisms which govern the charm production from pp to AA collisions.
Several nuclear effects can break the binary scaling estimated on a geometrical basis with the Glauber model~\cite{glauber}. They are divided into two classes: initial- and final-state effects. The former, such as nuclear shadowing, affect heavy-quark production by modifying the parton distribution functions in the nucleus. The latter can be due to the interactions of the partons in the medium. 
They affect heavy-quark production by modifying the fragmentation function.
Energy loss of the partons and development of anisotropic flow are among such processes.


\section{Measurements of open charm in the ALICE experiment}
A benchmark study was performed on the reconstruction of D$^0$ mesons in the $K^- \pi^+$ decay channel~\cite{PPR2cap6}. The study of ratios such as D$^+/$D$^0$, D$_s^+/$D$^0$, D$_s^+/$D$^+$ provide information on the hadronization mechanisms.
In addition, the reconstruction of as many charmed states as possible (D$^0$, D$^+$, D$_s^+$, $\Lambda_c$,\ldots) is required in order to reduce the systematic errors on the charm cross section.
The detection of charmed particles is possible in ALICE thanks to the good expected performance of the central barrel detectors, namely the Inner Tracking System (ITS), the Time Projection Chamber (TPC) and the Time Of Flight detector (TOF). The ALICE barrel is embedded in a magnetic field, B=0.5 T, oriented along the beam axis ($z$-axis). The tracks are curved in the transverse plane ($r\varphi$).
The expected $p_T$ resolution is $\sim 1 \%$, the primary vertex resolution is expected to be $\sim 10$ $\mu$m (in case of Pb-Pb) and the expected resolution on the minimum $r\varphi$ distance of the prolonged track with respect to the primary vertex (impact parameter) is $\sim 60$ $\mu$m for $p_T \simeq  1$ GeV/c. ALICE provides charged hadron identification from momenta of about 0.3 GeV/c to about 3 GeV/c.

\subsection{Reconstruction of $\DtoKpipi$ decay vertices}
An accurate measurement of secondary vertices is important because the weakly decaying charm states have typical decay lengths of the order of hundreds $\mu$m. 
The method for the reconstruction of secondary vertices originated by 3-prong decays  is based on the straight line approximation of the tracks, i.e. each track which is curved by the magnetic field is approximated by a straight line in the vicinity of the primary vertex\footnote{We verified that the error coming from the straight line approximation is negligible.}.
The algorithm finds the point of minimum distance between three straight lines by minimizing the quantity $D^2=\Sigma_{i=1,2,3}\frac{(x_i-x_0)^2}{\sigma^2_{xi}}+\frac{(y_i-y_0)^2}{\sigma^2_{yi}}+\frac{(z_i-z_0)^2}{\sigma^2_{zi}}$,
where $(x_0,y_0,z_0)$ are the coordinates of the secondary vertex and $\sigma_{x}$, $\sigma_{y}$, $\sigma_{z}$ are the errors on the track parameters.
At high $p_T$ of the D$^+$, its decay products get more and more collinear with the D meson $p_T$. Hence, along this direction a worsening of the resolution of the secondary vertex finder should be observed. This behaviour is illustrated in the left panel of Fig.~\ref{fig:vert}, where the RMS is calculated in a reference system in the bending plane in which $x$ is the direction of the $p_T$ of the D$^+$ and $y$ is the axis orthogonal to $x$. Along the $x$ coordinate the RMS increases at high $p_T$, while in the $y$ coordinate the RMS goes down to $\sim 40$ $\mu$m.
\section{Selection strategy for the $\DtoKpipi$ channel in Pb-Pb collisions.}
\begin{figure}[]
  \begin{center}
    \begin{tabular}{cc}
\includegraphics[width=.5\textwidth]{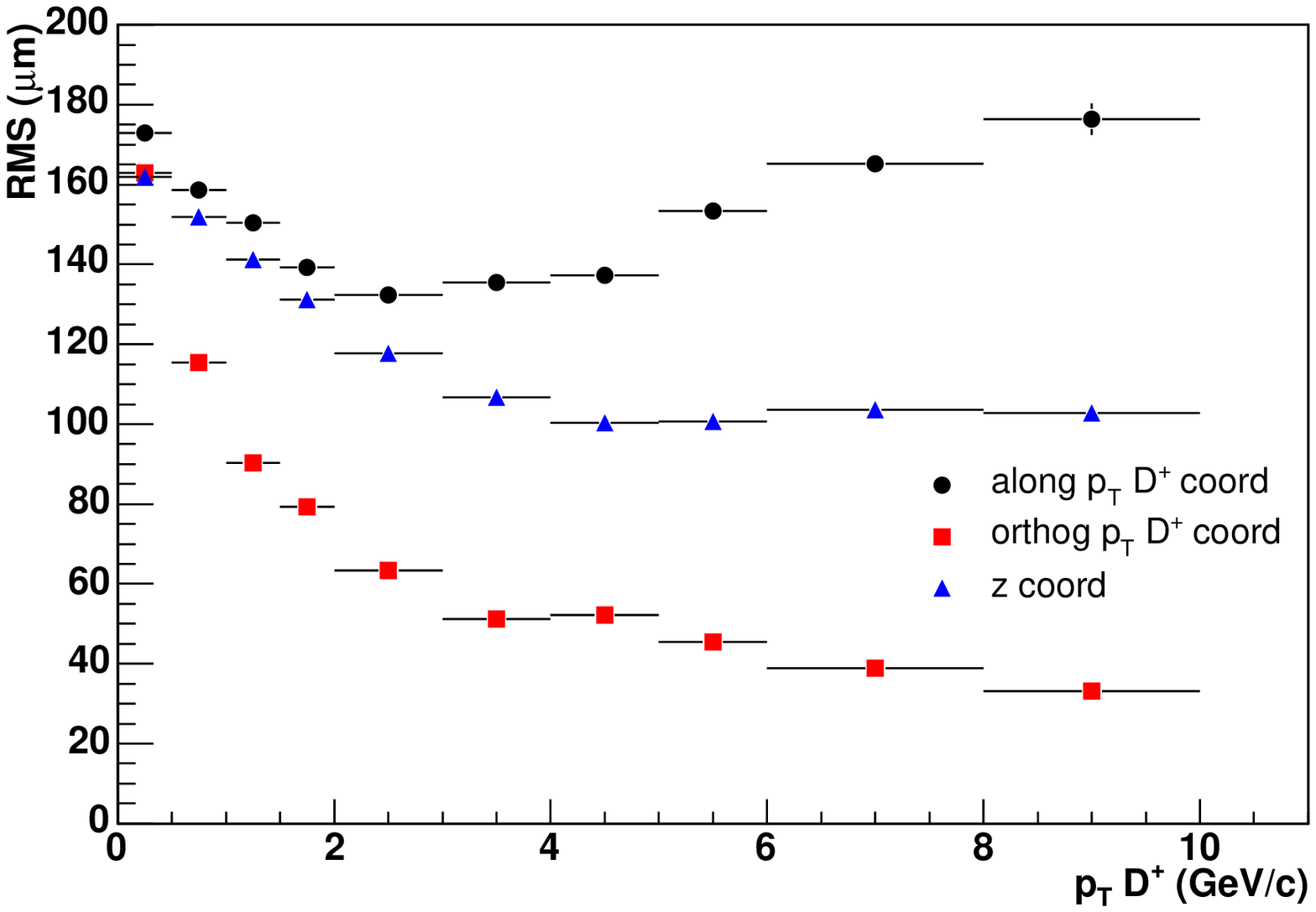}&     
 \includegraphics[width=0.48\textwidth]{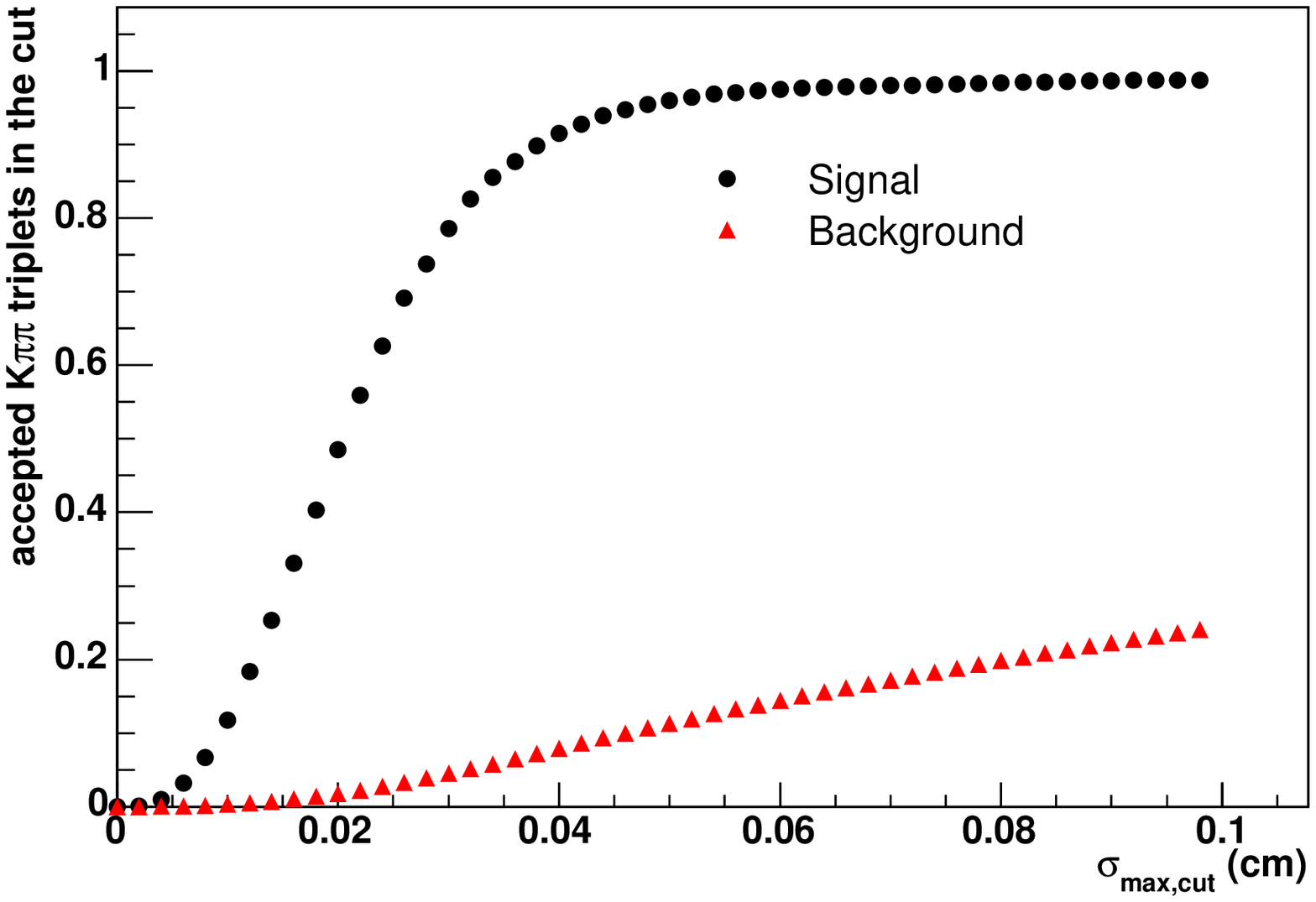} \\

 \end{tabular}
    \caption{Left panel: RMS of the residual distribution for the $x, y$ and $z$ coordinates of the secondary vertex as a function of the $p_T$ of the D$^+$. Right panel: ratio of triplets (both signal and background) selected by the cut $\sigma<\sigma_{cut}$, for D$^\pm$ transverse momentum $2<p_T<3$ GeV/c.}\label{fig:vert}
  \end{center}
\end{figure}

The expected number of D$^+$ in a central Pb-Pb event is $\sim 4$ in $|y|<1$~\cite{PPR2cap6}. We generated separately signal and background events. The results presented in this paper were obtained with a sample of 20~000 central HIJING (d$N_{ch}$/d$y$=6000) events and $\sim 2\times 10^6$ signal triplets generated with PYTHIA. The signal events have approximately the same particle multiplicity as the central HIJING events. A sample of $\sim 5.4\times 10^6$ minimum bias pp events was generated with PYTHIA; due to the smaller CPU time needed to simulate and reconstruct pp events compared to Pb-Pb, a separate generation of signal and background events was not necessary.
The cuts presented in the following are optimized in order to maximize the significance $S/\sqrt{S+B}$ in the invariant mass range $|M_{inv}-M_{D^{\pm}}|<1\sigma=12$~MeV/c$^2$.
The strategy is based on a series of selection steps.
The first selection is done on the single tracks 
before they are combined into triplets and passed to the secondary vertex finding algorithm. The goal is to reduce the number af all the possible combinations of tracks into triplets in a central Pb-Pb collision. This number is $\sim$ 10$^{10}$ without any initial cut and it must be cut before using the vertexer. The best combination of cuts which preserves the $p_T$ spectrum of D$^\pm$ mesons down to $\sim 1$ GeV/c is defined by $p_{Tcut,K}=0.7$ GeV/c, $p_{Tcut,\pi}=0.5$ GeV/c and $d_{0,cut}=95$ $\mu$m ($d_0$ is the distance of closest approach of the track to the primary vertex in the bending plane); with such a choice, the number of combinatorial background triplets per event is reduced to $\sim 10^6$.
The second selection takes place before the triplets are passed to the vertexer and it is based on a cut on the distance $\delta$ between the primary vertex and the secondary vertex of two opposite charged tracks.
 The chosen value of $\delta_{cut}$ is $\sim 700$ $\mu$m.
The triplets satisfying specific requirements on combinations of track impact parameters are then selected and the secondary vertex finding is applied.
A selection based on the track dispersion is performed ($\sigma^2=\Sigma_{i=1,2,3}(x_i-x_0)^2+(y_i-y_0)^2+(z_i-z_0)^2$\footnote{No weight with the errors on the track parameters was used; in this way $\sigma$ has the meaning of a standard deviation.}). The right-hand plot in Fig.~\ref{fig:vert} reports the fraction of accepted triplets (both signal and background) with $\sigma<\sigma_{cut}$, for D$^\pm$ transverse momentum $2<p_T<3$ GeV/c. 
The value $\sigma_{cut}$ has been tuned in various intervals of D$^\pm$ transverse momentum; it is of the order of 200 $\mu$m.
The candidate triplets undergo additional selections. The cut variables which have been considered at this level of the selection strategy are the following: 
\begin{enumerate}
\item \textit{Distance between the primary and the secondary vertices, d}. The signal triplets are characterized by larger values of $d$, according to the fact that they originate from displaced decay vertices
\item \textit{Maximum transverse momentum among the three tracks belonging to a triplet}, indicated as Max$\{p_{TK},p_{T\pi1},p_{T\pi2}\}$. This cut accounts for the slightly harder $p_T$ of the decay products (pions and kaons) of D$^\pm$ mesons compared to all the pions and kaons produced in a Pb-Pb event.

\item  $\cos \theta_{point}$, where $\theta_{point}$ is the pointing angle, defined by the direction of the reconstructed D$^\pm$ momentum on the bending plane and the line connecting the primary and the secondary vertices. If the found vertex really corresponds to a D$^\pm$ decay vertex, then $\theta_{point} \sim 0$ and $\cos \theta_{point} \sim 1$.
\item \textit{Sum of the squares of the impact parameters of the three tracks with respect to the primary vertex}, $s^2=\sum_{i=1,2,3} d_{0,i}^2=d_{0,K}^2+d_{0,\pi_1}^2+d_{0,\pi_2}^2$. This variable is expected to be larger for signal triplets which are displaced from the primary vertex.

\end{enumerate}
A method which simultaneously tunes all the four cut variables described above was developed. This method is based on the creation of multi-dimensional matrices for both signal and background events. Each cell $ijkl$ of the multi-dimensional matrices of signal and background contains the number of triplets which pass the selection criteria defined by the set of cuts: $C_{ijkl}=\{[d>d_{cut,i}] ~~AND~~[p_M>p_{Mcut,j}]~~AND~~[\cos\theta_{point}>\cos\theta_{point,cut~k}]~~AND~~[\Sigma d_{0,i}^2>(\Sigma d_{0,i}^2)_{cut,l}]\}$.
This procedure is applied in several intervals of D$^\pm$ transverse momentum $p_T$. Three analyses were performed in Pb-Pb, namely with ideal PID, with real PID and without any PID. The D$^\pm$ reconstruction was also studied in pp collisions assuming no particle identification.
\section{Results}
\begin{figure}[hbtp]
  \begin{center}
    \includegraphics[width=.5\textwidth]{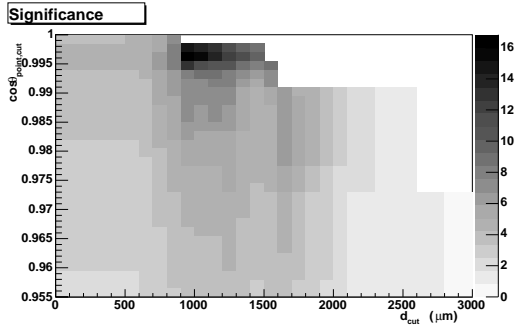}
    \caption{Significance (for 10$^7$ Pb-Pb central events) as a function of the cut variables $d_{cut}$ and cos$\theta_{point,cut}$, for $p_{M,cut}=0.5$ GeV/c and $(\Sigma d_{0,i}^2)_{cut}=0$ $\mu$m$^2$. The transverse momentum of the triplets is $2<p_T<3$ GeV/c. The real PID case is considered.} 
    \label{fig:signifxz_2PID}
  \end{center}
\end{figure}

\begin{figure}[hbtp]
  \begin{center}
    \begin{tabular}{cc}
  \includegraphics[width=.5\textwidth]{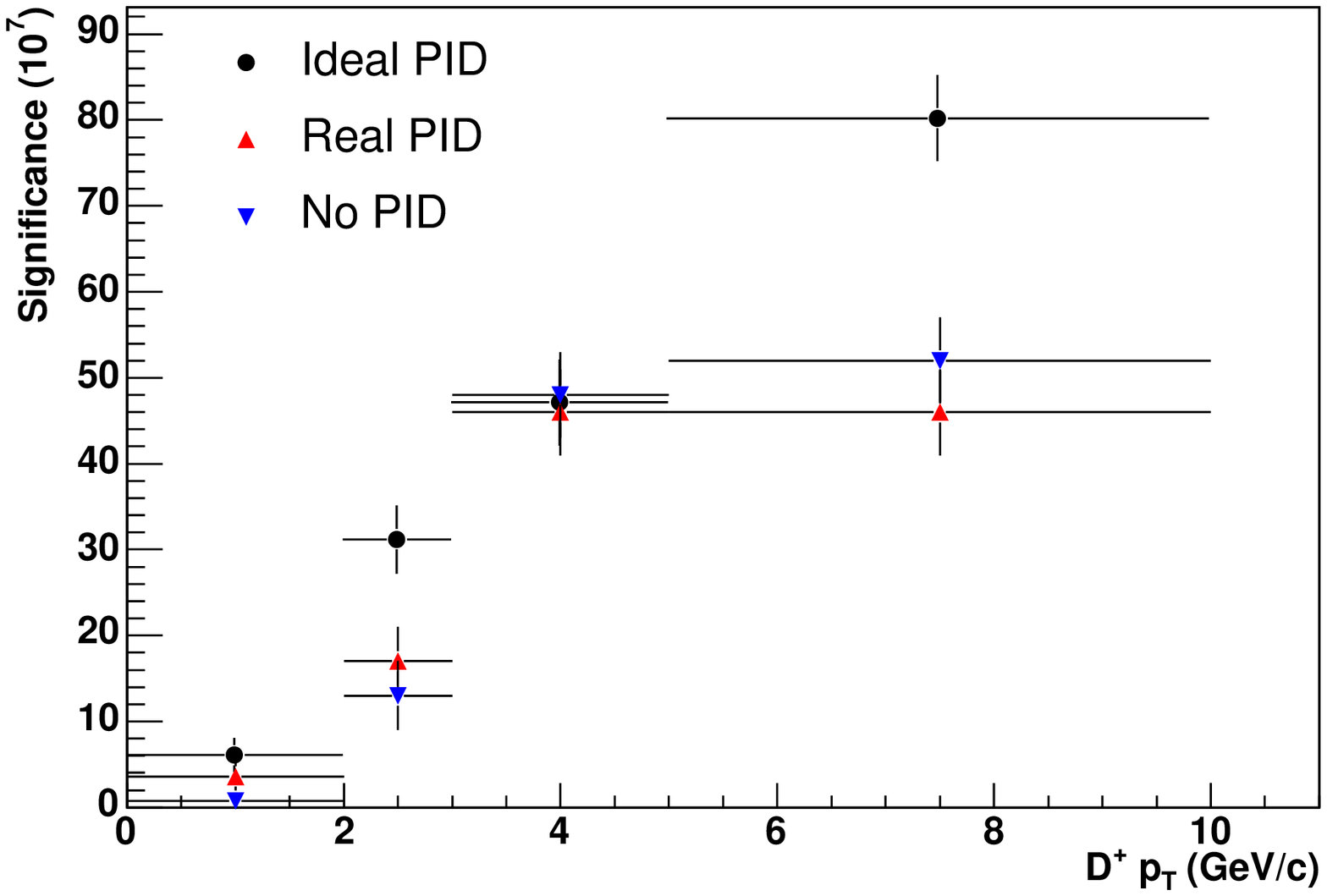}&
   \includegraphics[width=.5\textwidth]{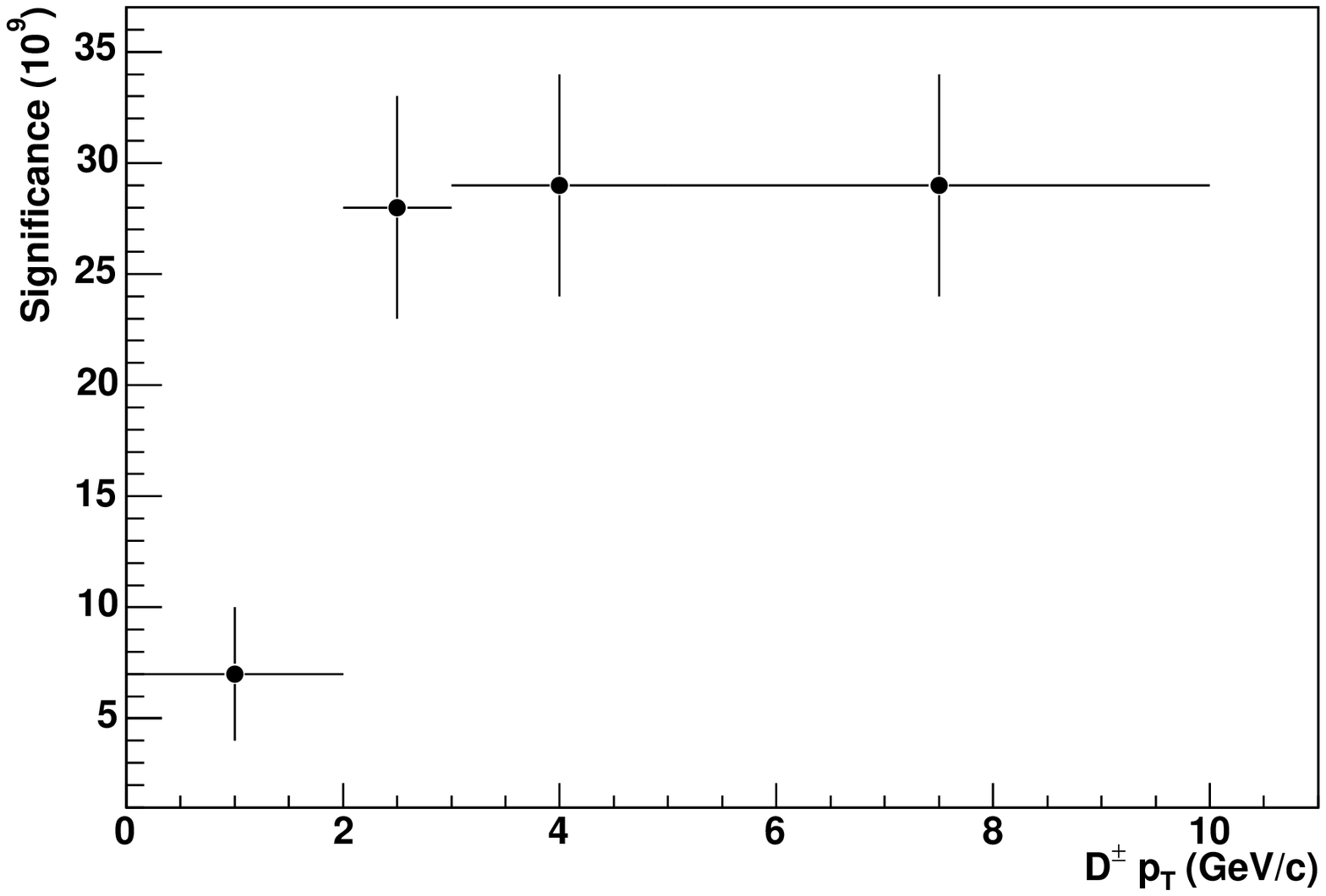}\\
 \end{tabular}
    \caption{Left panel: significance $S/\sqrt{S+B}$ (normalized to 10$^7$ central Pb-Pb events) as a function of the D$^\pm$ transverse momentum for ideal PID (black circles), real PID (red triangles up) and no PID (blue triangles down). Right panel: significance (normalized to 10$^9$ minimum bias pp collisions) as a function of the D$^\pm$ transverse momentum for pp collisions without PID.}\label{fig:SignificancevspT}
    
  \end{center}
\end{figure}

\begin{figure}[bhpt]
  \begin{center}
    \begin{tabular}{rl}
      \includegraphics[width=0.48\textwidth]{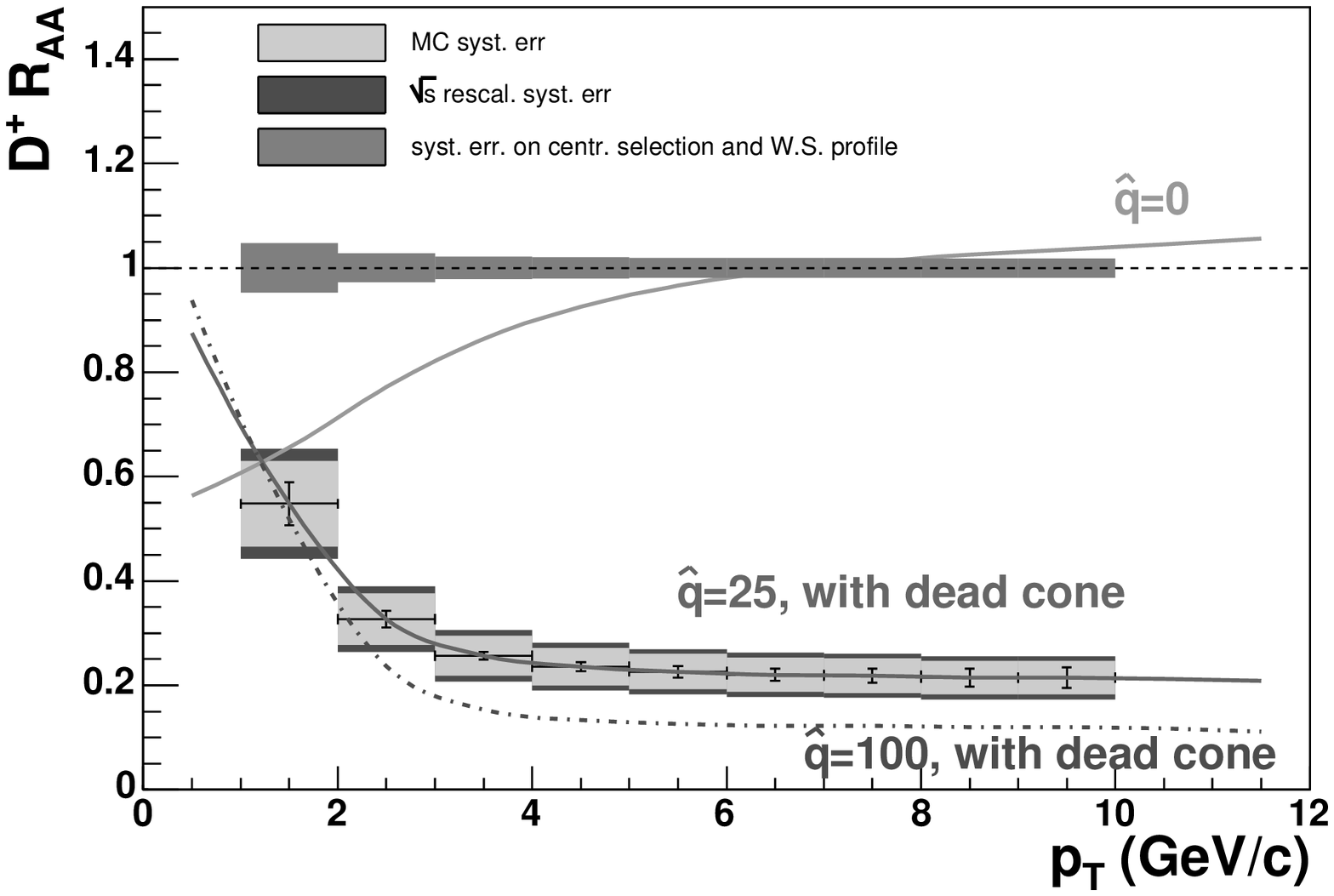}&
      \includegraphics[width=.51\textwidth]{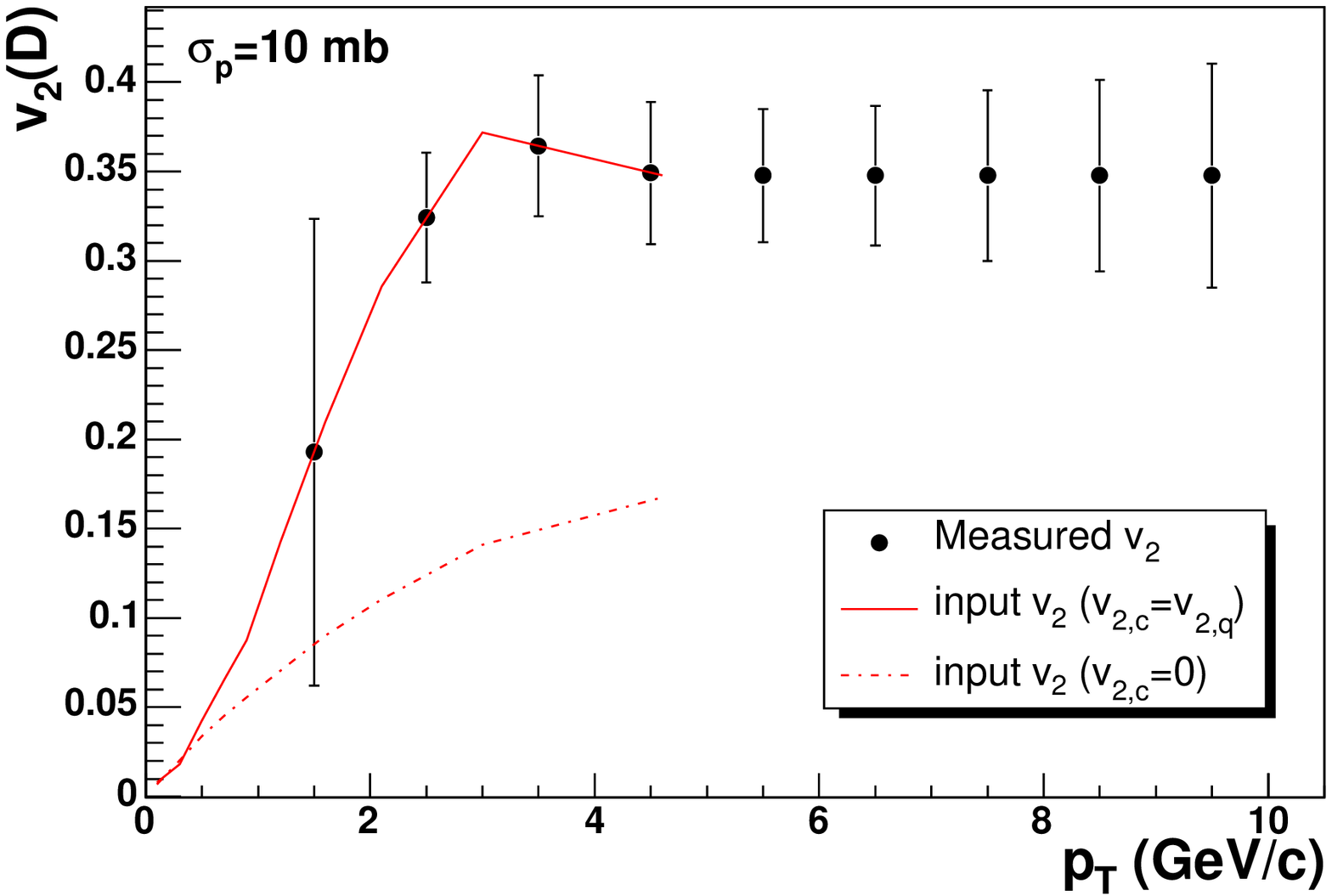}\\
    \end{tabular}
    \caption{Left panel: The $R_{AA}$ of D$^\pm$ mesons as a function of $p_T$, calculated with the parameter $\hat q=25$ GeV$^2$/fm (dark grey curve). The statistical and systematic errors are described in the legend. 
The results are compared to the scenario  without quenching of D$^\pm$ ($\hat q=0$ GeV$^2$/fm, light grey curve) and  with $\hat q=100$ GeV$^2$/fm including the dead cone effect (dashed-dotted light grey line). 
Right panel: D$^\pm$ meson $v_2^D(p_T)$ as a function of $p_T$ in the centrality class $6<b<9$ fm for a choice of the transport model parameter $\sigma_p=10$ mb. 
}\label{fig:persp}
  \end{center}
\end{figure}

An example of a result of the optimization of the cuts with the four-dimensional matrix procedure for $2<p_T<3$ GeV/c is reported in Fig.~\ref{fig:signifxz_2PID}, where the matrix of the significance is projected on the plane (cos$\theta_{point,cut}$, $d_{cut}$), for $(\Sigma d_{0,i}^2)_{cut}=0$ $\mu$m$^2$\footnote{This cut is not useful in this particular $p_T$ bin.} and $p_{M,cut}=0.5$ GeV/c.
The significance $S/\sqrt{S+B}$ maximized in the D$^\pm$ transverse momentum bins $0<p_T<2$ GeV/c, $2<p_T<3$ GeV/c, $3<p_T<5$ GeV/c, $p_T>5$ GeV/c is reported in Fig.~\ref{fig:SignificancevspT} for both Pb-Pb collisions (for the three cases of ideal PID, real PID and without PID) and pp collisions (without PID).
The obtained results were used to evaluate the statistical uncertainities related to the measurement of the nuclear modification factor\footnote{The nuclear modification factor is defined as: $R_{AA}^D=\frac{{\rm yield}^D_{AA}}{{\rm yield}^D_{pp}\times N_{coll}}$} and of the $v_2$ coefficient~\cite{voloshin}, which quantifies the amount of azimuthal anisotropy in the $p_T$ distribution. The errors are evaluated on a statistics of D$^\pm$ mesons expected for a one year run.
$R_{AA}$ is calculated~\cite{colcharge} as a function of $p_T$ in the case of no particle identification for both Pb-Pb and pp collisions and is shown in the left-hand panel of Fig.~\ref{fig:persp} for different values of the transport coefficient.
The performance for the measurement of the $v_2$ coefficient was estimated with a fast simulation consisting in the random generation of D$^\pm$ mesons distributed with an azimuthal anisotropy. The number of D$^\pm$ depends on $p_T$ and on centrality. The generated D$^\pm$ azimuthal angle is smeared with its experimental resolution.
Each of the generated D$^\pm$ is superimposed to an event made of tracks generated with a $v_2^{ev}$ coefficient (different from the D$^\pm$ meson $v_2^D$). These particles are used to reconstruct the event plane. The results of $v_2^D(p_T)$ for D$^\pm$ mesons are reported in Fig.~\ref{fig:persp} (right-hand panel); the values of $v_2^D$ of D$^\pm$ mesons used as input of the simulation are taken from~\cite{KO-LHC}. The predictions from~\cite{KO-LHC} are calculated in the framework of a coalescence model in case of charm flow (solid line) and in case of flow of only light quarks (dashed dotted line).

In conclusion, the reconstruction of the channel $\DtoKpipi$ seems to be feasible with a good significance both in Pb-Pb and in pp collisions. In the likely case in which the multiplicity in central Pb-Pb events turned out to be lower, the significance would of course improve. 
The analysis on $R_{AA}$ is feasible in one year of data taking. The statistical  error bars for the measurement of $v_2^D$ as a function of the transverse momentum in one year of data taking are a bit large. The use of a semi-peripheral trigger and the use of the D meson statistics collected in several years would improve the situation.
\medskip

I would like to thank Massimo Masera, Francesco Prino, Karel \v{S}afa\v{r}\'ik, Andrea Dainese and Federico Antinori for their fruitful suggestions.

\end{document}